\documentclass[conference]{IEEEtran}
\IEEEoverridecommandlockouts
\usepackage{cite}
\usepackage{amsmath,amssymb,amsfonts}
\usepackage{algorithmic}
\usepackage{graphicx}
\usepackage{textcomp}
\usepackage{xcolor}
\usepackage{caption}
\usepackage[left=1.62cm,right=1.62cm,top=1.9cm]{geometry}
\usepackage{subcaption}
\usepackage{graphicx}

\def\BibTeX{{\rm B\kern-.05em{\sc i\kern-.025em b}\kern-.08em
		T\kern-.1667em\lower.7ex\hbox{E}\kern-.125emX}}
\begin{document}
	
	\title{Quo Vadis, Optical Network Architecture? Towards an Optical-processing-enabled Paradigm}

	\author{\IEEEauthorblockN{Dao Thanh Hai}
		\IEEEauthorblockA{\textit{Posts and Telecommunications Institute of Technology} \\
			Hanoi, Vietnam \\
			haidt102@gmail.com}
	}
	
	\maketitle
	
	\begin{abstract} 
		In accommodating the explosive growth of Internet traffic, optical transport networks constituting the backbone of Internet infrastructure have been technologically and architecturally advancing. While technology-based advancements aim at expanding the system capacity by multi-order of magnitude, architectural solutions focus on optimal exploiting the available capacity by innovative ways of reducing the effective traffic load and therefore pave the way for serving more traffic requests. As a result, the evolution of optical network architectures is directed towards achieving the economic of scales by lowering the capital and operational expenditures per transmitted bit. Among various aspects, handling transit traffic at intermediate nodes represents a defining characteristic in classifying optical network architectures. In this context, the transition from the first generation of optical-electrical-optical (O-E-O) mode to the second generation of optical-bypass marked a paradigm shift in redesigning optical transport networks towards greater network efficiency. Optical-bypass operation has then become the \textit{de facto} approach adopted by the majority of carriers in both metro and backbone networks in the last two decades and has remained basically unchanged. However, in optical-bypass network, the fact that in-transit lightpaths crossing a common intermediate node must be separated in either time, frequency or spatial domain to avoid adversarial interference appears to be a critical shortcoming as the interaction of such lightpaths in optical domain may result in efficient computing and/or signal processing operations for saving spectral resources. Inspired by the accelerated progresses in optical signal processing technologies and the integration of computing and communications, we introduce in this paper a new architectural paradigm for future optical networks and highlight how this new architecture has the potential to shatter the \textit{status quo}. Indeed, our proposal is centered on exploiting the superposition of in-transit lightpaths at intermediate nodes to generate more spectrally efficient lightpaths and how to harness this opportunity from network design perspectives. We present two case studies featuring optical aggregation and optical XOR encoding to demonstrate the merit of optical-processing-enabled operation compared to its counterpart, optical-bypass. Numerical results on realistic network typologies are provided, revealing that a spectral saving up to $30\%$ could be achieved thanks to adopting optical-processing network.

	\end{abstract}
	
	
	\section{Introduction}
	Broadband Internet connections have become indispensable to our society, forming what is widely known as information era. The proliferation and popularization of bandwidth-intensive services enabled by Internet medium such as 4K and 8K ultra-high-definition video streaming, immersion into virtual reality (VR) and/or augmented reality (AR), and autonomous vehicles will certainly drive explosive growths of Internet traffic. As reported in \cite{futureoptics1, futureoptics2}, Internet traffic has been remaining an unabated growth annually of around $30\%$ and this means that in next ten years, roughly 14-fold larger the network capacity will be needed. Indeed, the currently deployed optical transport networks underpinning the majority of Internet traffic transmission will be facing capacity crunch under such rapid traffic rise and this necessitate for sustainably technological and architectural innovations to be investigated and then deployed to upgrade the network capacity \cite{futureoptics2, 20years}. While technology-based advancements aim at expanding the capacity by multi-order of magnitude, architectural solutions focus on optimal exploiting the available capacity by innovative ways of reducing the effective traffic load and therefore pave the way for serving more traffic requests. Major technological breakthroughs including coherent transmission, higher-order modulation formats, spectrally and spatially flexible optical transmission and recently wide-band optical transmission have been manifested in a recent world record data transmission set by NICT in 2021 featuring more than 300 Terabit/s across more than 3000 km \cite{NICT}. It has to be noted that increasing the system capacity alone does not result in appreciable improvements in economic of scales. Equally important, architectural solutions are on one hand to exploit better the system capacity and on the other hand, to achieve economic of scale by lowering both capital and operation expenditures per transmitted bit. Accordingly, optical network architecture has undergone a major transition in the 2000s time frame with the arrival of optical-bypass-enabled framework in replacing the costly optical-electrical-optical mode \cite{all-optical}. However, in optical-bypass networking, the fact that two (or more) lightpaths crossing intermediate nodes should be maximally separated to avoid interference which is considered as unwanted noise have turned out to be a fundamental limitation. Indeed, such long-established assumption perceiving the interference of optical channels transiting at the same node as a destructive factor and should therefore circumvent, albeit justifiable, may leave vastly unexplored opportunities. \\ 
	
	Different from electronic, photonic processing can make use of all four physical dimensions including amplitude, wavelength, phase and polarization to achieve a wide range of signal processing and computing capabilities. All-optical processing is envisioned therefore as the blueprint for future communication networks to overcome electronic bottleneck and to get in Tb/s regime in a super-efficient way \cite{all-optical, optical_processing_1}. Inspired by the accelerated progresses in optical signal processing technologies and the integration of computing and communications, we introduce in this paper a new architectural paradigm for future optical networks and highlight how this new architecture has the potential to shatter the \textit{status quo}. Indeed, our proposal is centered on exploiting the superposition of in-transit lightpaths at intermediate nodes to generate more spectrally efficient lightpaths and how to harness this opportunity from network design perspectives. In reversing the traditional assumption of keeping transitional optical channels intact as in optical-bypass networking, our proposal aims to re-define the optical node architecture by upgrading its naive functionalities from simply add/drop and cross-connecting to pro-actively combining transitional optical channels in photonic domain to better utilize spectrum capacity. We present two case studies featuring optical aggregation and optical XOR encoding to demonstrate the merit of optical-processing-enabled operation compared to its counterpart, optical-bypass. Numerical results on realistic network typologies are provided, revealing that a spectral saving up to more than $30\%$ could be achieved thanks to adopting optical-processing network. \\
	
	The paper is organized as followed. In Sect. II, the evolution of optical network architecture with respect to handling the transit traffic is examined, covering the historic optical-electrical-optical mode to the currently deployed optical-bypass to our proposal for what comes next, that is, optical-processing-enabled network. Besides, in order to highlight the potential of optical-processing architecture, two scenarios exploiting optical aggregation and optical XOR encoding are carefully analyzed. Section III is dedicated to report numerical results drawing on realistic topologies comparing spectral efficiency between conventional optical-bypass routing and our proposal. Finally, conclusions and forward-looking remarks are discussed in Sect. IV.

	\section{Evolution of Optical Network Architectures: From Optical-Electrical-Optical to Optical-bypass-enabled to Optical-processing-enabled Mode}
	
	Optical communications and networks have gone a long way in terms of capacity and efficiency since its birth roughly 50 years ago with Charles Kao's innovative proposal. It is important to note that fiber-optic communications not only provide a huge venue for data transmission but also give rise to optical networks and the field of optical networking. The development of optical networks is broadly linked to capability in manipulating lightstream. Specifically, handling transit traffic at intermediate nodes represents a defining characteristic in classifying optical network architectures \cite{hai_thesis}. \\
	
	The first generation of architecture was optical-electrical-optical (O-E-O) mode where a lightpath is terminated and regenerated at every intermediate node. A network configured in this mode performs O-E-O conversion of the signals at every endpoint of each transmission system. While having advantages such as eliminating cascaded physical impairments, allowing multi-vendor interoperability and easing performance monitoring, the downsize of this operation mode are manifold, notably with respect to very high cost of capital and operational expenditures \cite{hai_thesis}. Figure 1 illustrates how the transitional traffic is handled at each intermediate node by making use of an electrical/optical switch and a bank of transponders. 
	
	\begin{figure}[!htbp]
		\centering
		\includegraphics[width=\linewidth, height = 6cm]{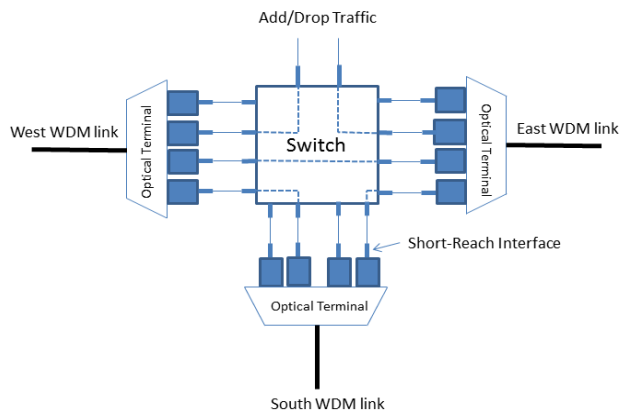}
		\caption{Optical-Electrical-Optical Node}
		\label{fig:opaque}
	\end{figure}

	Thanks to the convergence of several advances in optical transmission and switching technologies, in the year of 2000s, the introduction and then widespread adoption of optical-bypass mechanism where a lightpath remains fully in optical domain along its route marked the second generation in the development of optical network architecture. Figure 2 depicts the use of Reconfigurable Optical Add-Drop Multiplexer (ROADM) to perform optical cross-connect between input links and output links, permitting the remaining in optical domain of transit lightpaths. Optical-bypass operation has then become a dominant technology adopted by the majority of carriers in both metro and backbone networks in the last two decades \cite{all-optical}. In improving the optical-bypass architecture, there have been a number of proposal investigating the integration of simple optical signal processing functions including regeneration and  wavelength/format conversion for each individual in-transit lightpath. Nevertheless, the foundation of optical-bypass networking paradigm has remained basically unchanged that each lightpath crossing an intermediate node should be safeguarded from others in certain physical dimensions to mitigate unwanted interference which may degrade the optical signal qualities. \\

	\begin{figure}[!htbp]
		\centering
		\includegraphics[width=0.8\linewidth, height = 5cm]{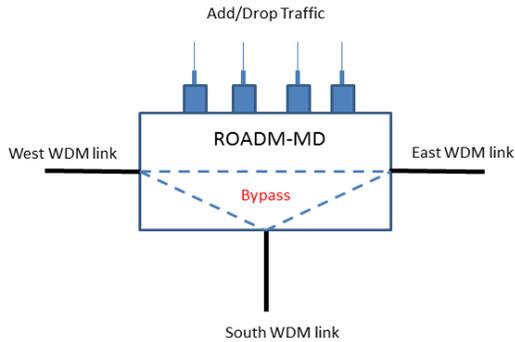}
		\caption{Optical-Bypass-enabled Node}
		\label{fig:bypass}
	\end{figure}

		\begin{figure}[!htbp]
		\centering
		\includegraphics[width=\linewidth, height = 5cm]{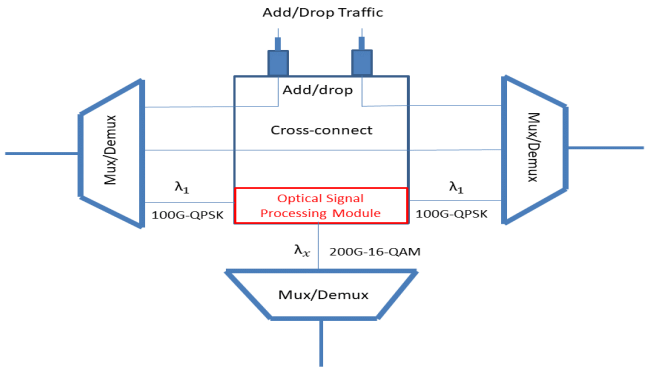}
		\caption{Optical-processing-enabled Node}
		\label{fig:processing}
	\end{figure}

	In reversing the well-established wisdom of treating in-transit lightpaths in optical-bypass, we propose a new architectural paradigm, entitled, optical-processing-enabled network, leveraging the opportunity of performing optical mixing between transitional lightpaths. This is enabled by the integration of optical signal processing modules to optical switch as conceptually shown in Fig. 3. Optical-processing-enabled paradigm is defined by the key property that optical nodes are empowered with optical processing functions. Specifically, two (or more) transiting lightpaths crossing an intermediate node could be optically superimposed to each other to generate the new output signal which is more spectrum-efficient than its inputs. We highlight the potential benefits of doing so by making use of two optical operations whose enabling technologies have been maturing in recent years, that is, optical aggregation and optical XOR \cite{ref1_22, ref2_22, ref3_22, optical_processing_2, hai_systems, hai_comcom, hai_comletter}. \\

	Figure 4 illustrates the routing, modulation selection and spectrum assignment for two demands with dedicated protection in the conventional optical-bypass framework. We turn the attention to Fig. 5  where optical node X is empowered with optical aggregation capability. Fig. 5(a) depicts an optical aggregator (de-aggregator) that receives two input signals of different bit-rate and modulation formats and generates an output signal whose bit-rate is equal to the sum of two inputs and is modulated at higher-order format. By permitting such operation at node X, instead of two optical channels $A_p$ and $B_p$ consuming a total of 11 spectrum slices, an aggregated lightpath 16-QAM carrying 300 Gbps and utilizing 6 spectrum slices is sufficient and this is highlighted in Fig. 5(b). In addition to the spectral savings, potential gains in operational expenditure involving energy consumption could be foreseen. Next we consider the second scenario where the optical XOR operation is armed at node X. The operation of such optical XOR encoder (decoder) is illustrated in Fig. 6(a). In this case, the protection channels of two demands are encoded together at node $X$ as a single 16-QAM signal carrying 200 Gbps (Fig. 6(b)). Rather than utilizing 11 spectrum slices on link $XE$ and $EZ$, 4 spectrum slices is enough to carry an encoded lightpath 200Gbps modulated on 16-QAM format. In case of single-link failure, the destination node $Z$ always receive two remaining signals permitting the reconstruct of lost signal by XOR operation as $(A_p \oplus B_p) \oplus A_p = B_p$ \\

	\begin{figure}[!htbp]
		\centering
		\includegraphics[width=\linewidth, height = 5cm]{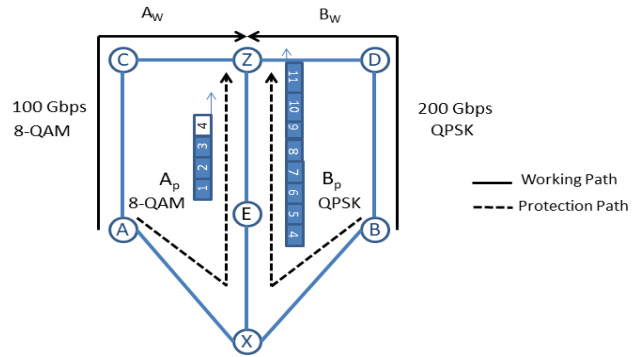}
		\caption{Traffic Provisioning in Optical-bypass Networks}
		\label{fig:test1}
	\end{figure}
		
	\begin{figure}[!htbp]
			\centering
			\includegraphics[width=\linewidth, height = 5cm]{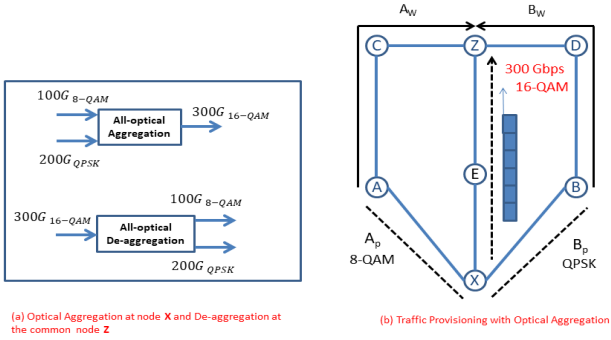}
			\caption{Traffic Provisioning with Optical Aggregation}
			\label{fig:test2}
	\end{figure}

	\begin{figure}[!htbp]
		\centering
			\includegraphics[width=\linewidth, height = 5cm]{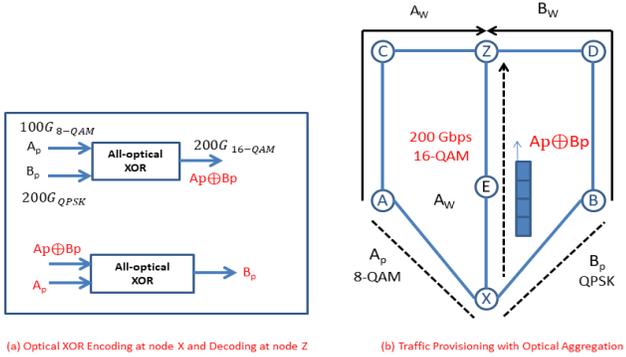}
			\caption{Traffic Provisioning with Optical Encoding}
			\label{fig:test3}
	\end{figure}

The aforementioned example highlights the potential benefits of exploiting the optical computing capabilities between in-transit lightpaths to achieve greater spectral efficiency. In the next part, we will report numerical results evaluating the impact of introducing such optical mixing operations to network designs. 

\section{Numerical Results }
This section is dedicated to provide a preliminary comparison between optical-processing-enabled and optical-bypass architecture in term of spectral efficiency when supporting the same set of traffic. Network designs for both architectures are formulated in the form of integer linear programming and optimal results are collected to guarantee a fair and reliable benchmark. Two cases studies featuring two optical mixing operations, that is, optical aggregation and optical XOR coding, are considered. The metric for comparison in both two studies is the wavelength link cost. 

\subsection{A Case Study of Optical-processing-enabled Network with Optical Aggregation}
In this part, we present a comparison between two designs based on optical-processing-enabled mode with optical aggregation and the traditional optical-bypass. The optical aggreation to be used in this study is a simple one aggregating two 100G QPSK signals into a single 16-QAM channel carrying 200G. The network topology for evaluation is a realistic COST239 one as shown in Fig. 7. The traffic under consideration is generated following the two-to-many model where two nodes are randomly selected as sources and seven remaining nodes are arbitrarily chosen as the destinations. Ten traffic sets are generated according to that model. \\

Figure 8 reveals the wavelengthh link cost of two designs at different traffic sets. It can be seen that the aggregation-aware design offers strictly greater performance in 8 out of 10 cases compared to the traditional design of optical-bypass. In term of relative gain, the design with optical aggregation could be up to more than $30\%$ spectrally more efficient than its counterpart in this studied case.

\begin{figure}[!ht]
	\centering
	\includegraphics[width=0.65\linewidth, height=5cm]{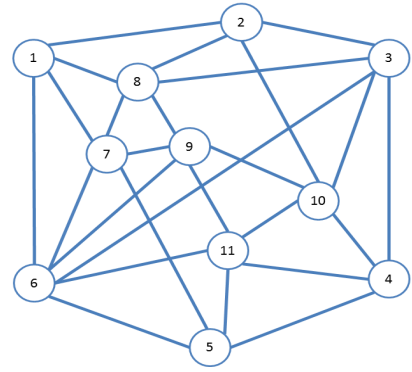}
	\caption{COST239 network}
	\label{fig:cost239}
\end{figure}

\begin{figure}[!ht]
	\centering
	\includegraphics[width=\linewidth, height=5.2cm]{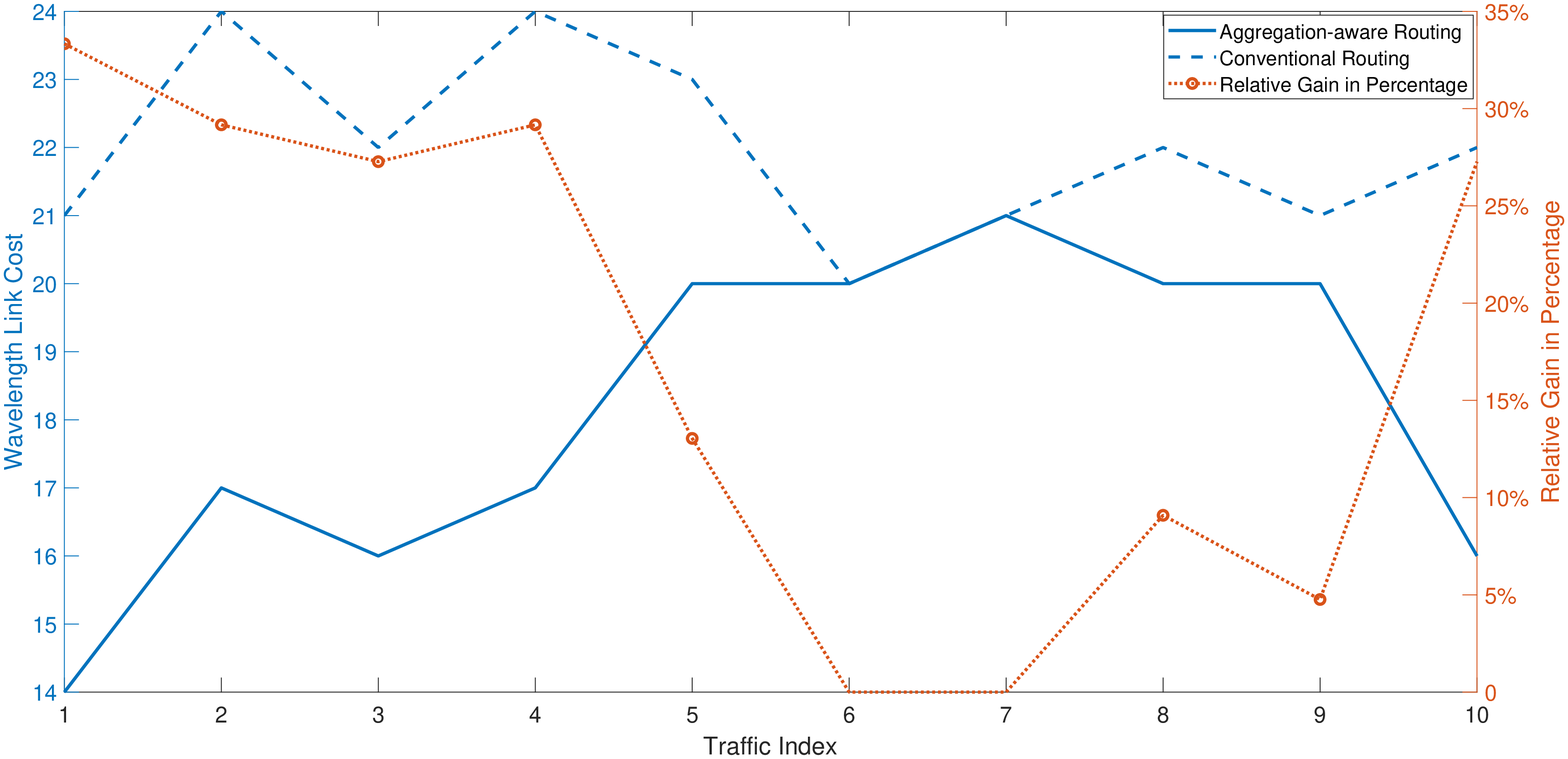}
	\caption{A Spectral Comparison between Aggregation-aware Routing in Optical-processing-enabled Network and Traditional One in Optical-bypass Network}
	\label{fig:aggregation}
\end{figure}

\subsection{A Case Study of Optical-processing-enabled Network with Optical XOR Encoding}
This part draws a numerical comparison between two designs based on optical-processing-enabled mode with optical XOR encoding and the traditional optical-bypass. The optical XOR encoding to be used in this study is a simple one performing XOR encoding between two optical signals of the same line rate and format. We restrict the study to a practical setting where the encoding is performed only on two demands with the same destination node and the decoding is only occurred at destination node. The network topology for evaluation is a realistic NSFNET one as shown in Fig. 9. To exploit network coding benefits, the traffic under consideration is generated following the two-to-many model where two nodes are randomly selected as sources and all the remaining nodes are designated as receiving ends. Ten random traffic sets are generated according to that model. \\

The results shown in Fig. 10 depicts the wavelength link cost of two designs in supporting the same set of traffic demands. As could be seen, the coding-aware design consistently outperforms the traditional approach with optical-bypass and in our studied case, the relative gain found could be up to $17\%$. 

\begin{figure}[!ht]
	\centering
	\includegraphics[width=0.9\linewidth, height=5cm]{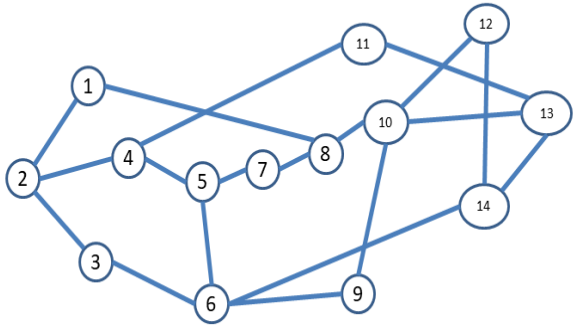}
	\caption{NSFNET network}
	\label{fig:nsfnet}
\end{figure}

\begin{figure}[!ht]
	\centering
	\includegraphics[width=\linewidth, height=5.2cm]{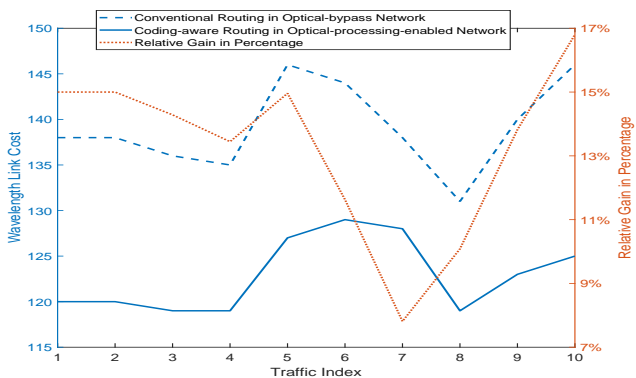}
	\caption{A Spectral Comparison between Coding-aware Routing in Optical-processing-enabled Network and Traditional One in Optical-bypass Network}
	\label{fig:coding}
\end{figure}

\section{Summary}	
We proposed the optical-processing-enabled architecture featuring the unique capability of pro-actively mixing transitional optical channels in photonic domain for future optical networks. The key idea is to enable multiple optical signals from in-transit lightpaths at different wavelengths and/or formats interacting with each other by non-linear processes in a controllable manner to achieve a desired output signal. In demonstrating the merit of our proposal, two case studies relying on optical aggregation and optical XOR coding were presented and it was shown that network design exploiting optical mixing operations could result in considerably greater performance than traditional optical-bypass approach. \\

En route from a visionary concept to practical deployments, many interesting and critical issue remain, motivating new technologies and innovations in the realm of optical communications and networking. The advent of fully optical-processing-enabled paradigm may then represent a tour de force for future transport networks. 



	\bibliographystyle{IEEEtran}
	\bibliography{ref}

\begin{thebibliography}{10}
\providecommand{\url}[1]{#1}
\csname url@samestyle\endcsname
\providecommand{\newblock}{\relax}
\providecommand{\bibinfo}[2]{#2}
\providecommand{\BIBentrySTDinterwordspacing}{\spaceskip=0pt\relax}
\providecommand{\BIBentryALTinterwordstretchfactor}{4}
\providecommand{\BIBentryALTinterwordspacing}{\spaceskip=\fontdimen2\font plus
\BIBentryALTinterwordstretchfactor\fontdimen3\font minus
  \fontdimen4\font\relax}
\providecommand{\BIBforeignlanguage}[2]{{%
\expandafter\ifx\csname l@#1\endcsname\relax
\typeout{** WARNING: IEEEtran.bst: No hyphenation pattern has been}%
\typeout{** loaded for the language `#1'. Using the pattern for}%
\typeout{** the default language instead.}%
\else
\language=\csname l@#1\endcsname
\fi
#2}}
\providecommand{\BIBdecl}{\relax}
\BIBdecl

\bibitem{futureoptics1}
A.~Lord \emph{et~al.}, ``Future optical networks in a 10 year time frame,'' in
  \emph{2021 Optical Fiber Communications Conference and Exhibition (OFC)},
  2021, pp. 1--3.

\bibitem{futureoptics2}
P.~J. Winzer, ``Capacity scaling through spatial parallelism: From subsea
  cables to short-reach optical links,'' in \emph{2021 Optical Fiber
  Communications Conference and Exhibition (OFC)}, 2021, pp. 1--1.

\bibitem{20years}
\BIBentryALTinterwordspacing
P.~J. Winzer \emph{et~al.}, ``Fiber-optic transmission and networking: the
  previous 20 and the next 20 years,'' \emph{Opt. Express}, vol.~26, no.~18,
  pp. 24\,190--24\,239, Sep 2018. [Online]. Available:
  \url{http://www.opticsexpress.org/abstract.cfm?URI=oe-26-18-24190}
\BIBentrySTDinterwordspacing

\bibitem{NICT}
\BIBentryALTinterwordspacing
(2021) Demonstration of world record: 319 tb/s transmission over 3,001 km with
  4-core optical fiber. [Online]. Available:
  \url{https://www.nict.go.jp/en/press/2021/07/12-1.html}
\BIBentrySTDinterwordspacing

\bibitem{all-optical}
A.~Saleh and J.~M. Simmons, ``All-optical networking: Evolution, benefits,
  challenges, and future vision,'' \emph{Proceedings of the IEEE}, vol. 100,
  no.~5, pp. 1105--1117, May 2012.

\bibitem{optical_processing_1}
A.~E. Willner \emph{et~al.}, ``All-optical signal processing techniques for
  flexible networks,'' \emph{Journal of Lightwave Technology}, vol.~37, no.~1,
  pp. 21--35, 2019.

\bibitem{hai_thesis}
\BIBentryALTinterwordspacing
H.~DAO~THANH, ``{Contribution to Flexible Optical Network Design: Spectrum
  Assignment and Protection},'' Theses, {T{\'e}l{\'e}com Bretagne ;
  Universit{\'e} de Bretagne Occidentale}, Mar. 2014. [Online]. Available:
  \url{https://hal.archives-ouvertes.fr/tel-01206788}
\BIBentrySTDinterwordspacing

\bibitem{ref1_22}
\BIBentryALTinterwordspacing
M.~Rapisarda \emph{et~al.}, ``All-optical aggregation and distribution of
  traffic in large metropolitan area networks using multi-tb/s s-bvts,''
  \emph{J. Opt. Commun. Netw.}, vol.~14, no.~5, pp. 316--326, May 2022.
  [Online]. Available:
  \url{http://opg.optica.org/jocn/abstract.cfm?URI=jocn-14-5-316}
\BIBentrySTDinterwordspacing

\bibitem{ref2_22}
\BIBentryALTinterwordspacing
Y.~Ding \emph{et~al.}, ``Experimental demonstration of all-optical aggregation
  and de-aggregation for a qpsk signal in an elastic optical network,''
  \emph{Opt. Express}, vol.~30, no.~5, pp. 6456--6468, Feb 2022. [Online].
  Available: \url{http://opg.optica.org/oe/abstract.cfm?URI=oe-30-5-6456}
\BIBentrySTDinterwordspacing

\bibitem{ref3_22}
A.~Sueyoshi \emph{et~al.}, ``Multi-stage adaptive equalization for
  all-optical-aggregated 16qam signal,'' \emph{IEICE Communications Express},
  vol. advpub, 2022.

\bibitem{optical_processing_2}
A.~Fallahpour \emph{et~al.}, ``Demonstration of tunable optical aggregation of
  qpsk to 16-qam over optically generated nyquist pulse trains using nonlinear
  wave mixing and a kerr frequency comb,'' \emph{Journal of Lightwave
  Technology}, vol.~38, no.~2, pp. 359--365, 2020.

\bibitem{hai_systems}
D.~T. Hai \emph{et~al.}, ``A priority-based multiobjective design for routing,
  spectrum, and network coding assignment problem in network-coding-enabled
  elastic optical networks,'' \emph{IEEE Systems Journal}, vol.~14, no.~2, pp.
  2358--2369, 2020.

\bibitem{hai_comcom}
D.~T. Hai, ``A bi-objective integer linear programming model for the routing
  and network coding assignment problem in wdm optical networks with dedicated
  protection,'' \emph{Computer Communications}, vol. 133, pp. 51 -- 58, 2019.

\bibitem{hai_comletter}
------, ``Leveraging the survivable all-optical wdm network design with network
  coding assignment,'' \emph{IEEE Communications Letters}, vol.~21, no.~10, pp.
  2190--2193, Oct 2017.

\end{thebibliography}
	
	\vspace{12pt}
	\color{red}
	
\end{document}